**Babcock Redux: An Amendment of Babcock's Schematic of the Sun's Magnetic Cycle**

Ronald L. Moore[1,2], Jonathan W. Cirtain[1], and Alphonse C. Sterling[1]

[1] Heliophysics and Planetary Science Office, ZP13, Marshall Space Flight Center, AL 35812, USA;
ron.moore@nasa.gov

[2] Center for Space Plasma and Aeronomic Research, University of Alabama in Huntsville, AL 35899, USA

**Abstract**

We amend Babcock's original scenario for the global dynamo process that sustains the Sun's 22-year magnetic cycle. The amended scenario fits post-Babcock observed features of the magnetic activity cycle and convection zone, and is based on ideas of Spruit & Roberts (1983) about magnetic flux tubes in the convection zone. A sequence of four schematic cartoons lays out the proposed evolution of the global configuration of the magnetic field above, in, and at the bottom of the convection zone through sunspot Cycle 23 and into Cycle 24. Three key elements of the amended scenario are: (1) as the net following-polarity magnetic field from the sunspot-region $\Omega$-loop fields of an ongoing sunspot cycle is swept poleward to cancel and replace the opposite-polarity polar-cap field from the previous sunspot cycle, it remains connected to the ongoing sunspot cycle's toroidal source-field band at the bottom of the convection zone; (2) topological pumping by the convection zone's free convection keeps the horizontal extent of the poleward-migrating following-polarity field pushed to the bottom, forcing it to gradually cancel and replace old horizontal field below it that connects the ongoing-cycle source-field band to the previous-cycle polar-cap field; (3) in each polar hemisphere, by continually shearing the poloidal component of the settling new horizontal field, the latitudinal differential rotation low in the convection zone generates the next-cycle source-field band poleward of the ongoing-cycle band. The amended scenario is a more-plausible version of Babcock's scenario, and its viability can be explored by appropriate kinematic flux-transport solar-dynamo simulations.

1. INTRODUCTION

All of the magnetic field on the Sun evidently comes from Ω-loop field stitches that bubble up through the photosphere from below (Zwaan 1987). The evolving field in and above the photosphere modulates the Sun's luminosity, results in the mega-Kelvin hot corona and its solar-wind outflow, and explosively drives coronal mass ejections, flares, and myriad jets. The dynamo process, the generation of the source field inside the Sun, thus gives rise to all solar magnetic activity and consequent space weather and space climate. For this reason, and because even the rudiments of how the Sun's dynamo process works and how it endures remain uncertain and controversial (e.g., Charbonneau 2010; Spruit 2011), the solar dynamo problem is one of the grand challenges of solar astrophysics.

In this paper, the Sun's "external field" is the field in and above the photosphere. We take the "large-scale" external field to be the field that reaches heights above the photosphere greater than 0.05 $R_{Sun}$ and/or spans horizontal distances greater than 0.05 $R_{Sun}$, i.e., more than the diameter of a typical supergranule (~ 35,000 km). Except occasionally during a few solar rotations in the minimum phase of the solar cycle, much of the large-scale external field is in the form of active regions. [The "solar cycle" is the Sun's 11-year cycle of magnetic activity accompanying sunspots (Howard 1977). An "active region" is a large-scale Ω loop or cluster of large-scale Ω loops that has emerged recently enough that it can still be recognized as an entity in full-disk chromospheric and coronal images and full-disk magnetograms. When an active-region Ω-loop cluster has completed its emergence, it typically has one or more sunspots in each foot.] The rest of the large-scale field, the large-scale field not in active regions, is rooted in the magnetic network at the edges of supergranules and fills the corona in quiet regions and coronal holes (Zwaan 1987). It is made of merged dispersed remnants of decayed active regions that can no longer be individually discerned.

This paper offers a heuristic picture of the global dynamo process that generates the Sun's internal field from which the large-scale external field emerges. The proposed picture is a plausible amendment of the original solar dynamo scenario of Babcock (1961). Our intent is to provide for the solar dynamo a heuristic visualization that is of the style of Babcock's depiction, and stems from his scenario, but has a more plausible global arrangement and progression of the internal field from which the large-scale external field emerges. Our scenario is similar to Babcock's in that it graphically lays out the proposed three-dimensional global form and progression of the internal field. Our scenario is an improvement over Babcock's in having a more plausible global configuration of the internal field that is present late in each sunspot cycle and gives rise to the next sunspot cycle, and in having a clearer vision of how that internal field configuration could come naturally from the ongoing sunspot cycle. We are putting forth our scenario as a plausible idea, as a hypothesis worth considering. The degree to which it captures the essence of the Sun's actual global dynamo process is an open question to be addressed by equations-based magneto-convection simulations and equations-based dynamo simulations beyond the scope of this paper.

*1.1. Observed Solar-Cycle Features Known to Babcock*

For a year or two around the minimum in sunspot number between each 11-year sunspot cycle and the next, sunspots appear in two nested north-south pairs of latitude bands. The inner pair closely brackets the equator, each band centered less than 10° from the equator; each band of the outer pair is centered about 30° from the equator (e.g., Hathaway & Rightmire 2011). The sunspots near the equator are the last sunspots of the ending cycle and the sunspots in the higher-latitude bands are the first sunspots of the new cycle. As the new cycle progresses, the previous cycle ends, sunspots occur only in the two higher-latitude bands, and these bands gradually drift closer to the equator over the rest of the cycle. They become the low-latitude pair at the end of the cycle as next-cycle sunspots start appearing in the two sunspot bands of the next cycle, centered at ± 30° latitude. The equatorward migration of the two sunspot bands of each 11-year cycle is called Sporer's law (Zirin 1988).

When an active-region $\Omega$-loop cluster has completed its emergence, it is usually aligned roughly east-west, one foot leading the other in the direction of solar rotation. Throughout each 11-year sunspot cycle, in each band of that cycle's pair of equatorward-drifting bands of sunspot active regions, the east-west direction of the field in all but a small fraction of the active regions is the same, and is the opposite of that in the other band. Also, the field direction of each band is the opposite of that which the corresponding band has throughout the next cycle. These magnetic-polarity rules of the sunspot cycle are called the Hale-Nicholson law (Zirin 1988). The Hale-Nicholson law establishes that the Sun has a 22-year global magnetic cycle.

The sunspot bands, Sporer's law, and the Hale-Nicholson law together suggest that, during each sunspot cycle's long middle interval during which only one pair of sunspot bands is present, the Sun's large-scale field comes, by emergence of large-scale $\Omega$ loops, from a corresponding pair of bands of oppositely-directed internal toroidal field generated by the global dynamo process. To be the source of the sunspot-making $\Omega$ loops, each internal toroidal field band must track the equatorward migration of the corresponding sunspot band throughout each 11-year sunspot cycle and have the opposite field direction throughout the next cycle. We mentioned above that as the old cycle ends and the new cycle begins, new-cycle sunspots occur in the new cycle's sunspot bands while old-cycle sunspots are still occurring in the old cycle's sunspot bands. This suggests that during this transition time there are two oppositely-directed bands of internal toroidal field on each side of the equator, one centered about 30° from the equator and the other centered much closer to the equator.

When an active-region $\Omega$ loop has completed its emergence, it is seldom directed exactly east-west, but is usually discernibly tilted either toward the equator (leading foot at lower latitude than the trailing foot) or away from the equator (leading foot at higher latitude than the trailing foot). The average tilt of the sunspot-active-region $\Omega$ loops in each sunspot band is toward the equator. The average tilt decreases as the band moves closer to the equator, the average tilt angle decreasing from about 10° at the start of a sunspot cycle to about 3° at the end. This average behavior is called Joy's law (Zirin 1988).

*1.2 Synopsis of Babcock's Dynamo Scenario*

When Horace Babcock published his seminal schematic model of the Sun's global dynamo process (Babcock 1961), the Sun's magnetic field had been globally mapped and monitored by a disk-scanning magnetograph for only 9 years, covering less than an entire 11-year sunspot cycle (Babcock & Babcock 1995; Babcock 1959). These observations revealed that (1) from 1952, late in the decline of sunspot Cycle 18, until the end of 1956, late in the rise of Cycle 19, the large-scale field in each polar cap, the zone within about 30° of the pole, was unipolar and had the polarity of the trailing feet of the large-scale $\Omega$ loops of Cycle 18 in that hemisphere, and (2) during the 1957-1958 maximum phase of Cycle 19, the field in each polar cap reversed, was replaced by opposite-polarity field of about the same strength. These discoveries led Babcock to construct his solar dynamo scenario. A premise of his scenario is that during the maximum phase of each sunspot cycle there is a polar-cap field reversal similar to that in Cycle 19, each cap's field polarity changing from that of the trailing feet of the active-region $\Omega$ loops of the previous cycle in that hemisphere to that of the trailing feet of the active-region $\Omega$ loops of the ongoing cycle in that hemisphere. That premise has turned out to be true (Hathaway 2010).

In Babcock's scenario for the Sun's global dynamo process, about 3 years before the first high-latitude active-region $\Omega$ loops of a new sunspot cycle start emerging, the Sun's large-scale internal field is a purely poloidal continuation of the unipolar large-scale field in each polar cap. At that time, the field polarity in each cap is that of the following feet of the active-region $\Omega$ loops of the ongoing cycle in that hemisphere. At that time in Babcock's scenario, the large-scale internal field enters the negative-polarity cap, runs to the positive-polarity cap horizontally north-south at a depth of about 0.1 $R_{Sun}$ in the convection zone, and exits the positive-polarity cap. Babcock assumed that the Sun's latitudinal differential rotation, the increase in the rotation period of the surface from 25 days at the equator to 36 days at the poles (e.g., Komm et al 1993; Hathaway & Rightmire 2011), persists undiminished with depth to at least the depth of the internal field. In each polar hemisphere (north and south of the equator), the latitudinal differential rotation of the (high-beta) plasma in which the internal field resides, by continually shearing the poloidal component of the (initially purely poloidal) internal field, then builds a band of increasingly stronger internal toroidal field between the polar cap and the equator. Babcock argued that in about three years, the toroidal field becomes strong enough for large-scale $\Omega$ loops to arise from it and emerge through the photosphere, making sunspot active regions. Due to the Joy's-law tilt of the active regions, as the active-region fields spread out as they decay, more leading-polarity flux than following-polarity flux from each polar hemisphere meets and cancels opposite-polarity flux spreading from the other hemisphere. This leaves a net surplus of following-polarity flux in each polar hemisphere. Babcock assumed that on the poleward side of each sunspot band there is poleward flow of the top of the convection zone, flow that carries the net following-polarity flux to the polar cap. In the maximum phase of the ongoing sunspot cycle, the encroaching net following-polarity flux of the ongoing sunspot cycle gradually cancels out and replaces the opposite-polarity polar-cap flux (which came from the net following-polarity flux of the previous sunspot cycle). In Babcock's scenario, by late in the ongoing sunspot cycle (3 years before the onset of the next sunspot cycle) the Sun's global-scale internal field is again purely poloidal, as at the start of the ongoing sunspot cycle, but opposite in direction, ready for the generation of the toroidal field of the next sunspot cycle.

Babcock's scenario does not address how the purely poloidal internal field that the spawns the next sunspot cycle is put in place in the convection zone by late in each ongoing sunspot cycle. Neither does it explain the observed high-latitude start of the sunspot band of the next cycle during the low-latitude end of the sunspot band of the ongoing cycle. That is, Babcock's scenario ignores the presence of the internal toroidal field band of the ongoing sunspot cycle during the building and early active-region $\Omega$-loop production of the toroidal field band of the next sunspot cycle. The present paper aims to amend these two shortcomings of Babcock's scenario. We give a schematic visualization (cartoons) of how the internal toroidal field of the next sunspot cycle might be made during the ongoing sunspot cycle. Our scenario adopts two main ideas of Babcock's scenario: (1) the toroidal field band of each sunspot cycle is made from internal poloidal field from the previous sunspot cycle by the latitudinal differential rotation of the convection zone, and (2) Joy's law, via cancellation of leading-polarity flux across the equator, results in a net surplus of following-polarity flux in each hemisphere that reverses the polar-cap field during the maximum phase of each sunspot cycle.

## 2. POST-BABCOCK OBSERVATIONS

### 2.1. Torsional-Oscillation Wave and Extended Solar Activity Cycle

It is now known that each sunspot band is centered on and tracks the descending latitude of the peak in latitudinal differential-rotation shear in the so-called torsional-oscillation wave that propagates equatorward with the sunspot band (LaBonte & Howard 1982; Hathaway & Rightmire 2011; McIntosh et al 2014). The wave is a perturbation of a few meters per second amplitude in the Sun's differential rotation, with a wavelength of roughly the north-south span of the corresponding sunspot band (20°-40° of latitude). The wave's positive perturbation to the shear of the Sun's latitudinal differential rotation peaks midway between the lower-latitude faster-than-average-rotation half of the wave and the higher-latitude slower-than-average-rotation half.

The torsional-oscillation wave that tracks a sunspot band through a sunspot cycle originates earlier and at higher latitude than the first sunspots of the sunspot band. The wave begins during the maximum phase of the previous sunspot cycle and is initially centered about 60° from the equator. It then steadily drifts to the equator over the next 18-22 years in phase with the so-called extended solar activity cycle (Wilson et al 1988). During the 6-7 years before the center of the wave reaches 30° from the equator and sunspots start occurring in the corresponding sunspot band, sunspotless ephemeral active regions that preferentially have the east-west polarity direction of the coming sunspot band occur at high latitudes in a broad band of latitude (~30° wide) roughly centered on the wave (Wilson et al 1988). Also, enhanced coronal emission observed above the limb roughly centers on the wave and tracks the wave's equatorward progression (Wilson et al 1988). From analysis of the spatial distribution and separation of coronal EUV bright points (many of which are signatures of ephemeral active regions) and their evolving global pattern over the solar cycle, McIntosh et al (2014) have recently corroborated the above equatorward progression of the high-latitude ephemeral-region emergence and coronal brightening, and that it traces the equatorward progression of the torsional-oscillation wave and corresponding sunspot band.

As Wilson et al (1988) and McIntosh et al (2014) have each emphasized, the above observed aspects of the torsional-oscillation wave and extended solar activity cycle have two implications for the toroidal source-field band for each sunspot band.  One implication is that the toroidal source-field band originates together with the torsional-oscillation wave at polar latitudes near 60° during the maximum phase of the previous sunspot cycle and drifts with the wave to the equator.  The other implication is that the toroidal field band is the source of only ephemeral active regions when it is at latitudes above about 30°, and is the source of sunspot active regions as well when it is at lower latitudes.

*2.2. Rotation and Meridional Flow in the Convection Zone*

Helioseismic measurements of the Sun's interior have established that the interface, the so-called tachocline, between the bottom of the convection zone and the top of the radiative interior is centered at about 0.7 $R_{Sun}$, and that below about 0.65 $R_{Sun}$ down to at least 0.2 $R_{Sun}$ the radiative interior rotates as a rigid body and has a rotation period of 27 days (Howe 2009).  Helioseismic measurements of the convection zone have confirmed the latitudinal differential rotation of the top of the convection zone and the torsional-oscillation wave in the differential rotation of the top of the convection zone (Howe 2009).  Helioseismology has also revealed that the latitudinal differential rotation persists with depth in the convection zone, remaining largely unchanged down to about 0.75 $R_{Sun}$, and then rapidly transitions though the tachocline (from 0.75 $R_{Sun}$ down to 0.65 $R_{Sun}$) to the rigid-body rotation of the radiative interior (Howe 2009).  So, it is now known that there is latitudinal differential rotation at the bottom of the convection zone (that is, at the top of the tachocline) comparable to that at the top of the convection zone.  It can generate toriodial magnetic field there by shearing any horizontal poloidal field that happens to be at the bottom of the convection zone.

The so-called magnetic butterfly diagram (e.g., Hathaway 2010) is the time-latitude plot of the per-solar-rotation net magnetic flux density measured in small steps of latitude from pole to pole from a continuous series of full-disk magnetograms spanning a solar cycle or more.  It shows that throughout the 10°-60° latitude range in both polar hemispheres magnetic flux continually drifts poleward at a speed of ~10 m s$^{-1}$, revealing that in this latitude range there is persistent global poleward meridional flow of this speed in the top of the convection zone in and below the photosphere (Howard & LaBonte 1981; Topka et al 1982).  The poleward meridional flow of the top of the convection zone has been confirmed by correlation tracking of the magnetic network in full-disk magnetograms from the Helioseismic and Magnetic Imager (HMI) (Hathaway & Rightmire 2011), by correlation tracking of supergranule convection cell tops in the photosphere in HMI full-disk Dopplergrams (Hathaway 2012b), and by helioseismic time-distance analysis of HMI Doppler data (Zhao et al 2013).

Correlation tracking of supergranules in HMI full-disk Dopplergrams has also revealed that the meridional flow reverses direction from poleward to equatorward at a depth between 50,000 km and 70,000 km below the surface, that is, between 0.93 $R_{Sun}$ and 0.90 $R_{Sun}$ (Hathaway 2012b). From helioseismic time-distance analysis of HMI Doppler data, Zhao et al (2013) confirmed that the meridional flow reverses at this depth.  Their analysis found the merional flow to be equatorward at a speed of about 10 m s$^{-1}$ between 0.91 $R_{Sun}$ and 0.82 $R_{Sun}$, and to again reverse direction, from equatorward to poleward, with increasing depth at 0.82 $R_{Sun}$.  From helioseismic time-distance analysis of Doppler data from Global Oscillation Network Group (GONG), Jackiewicz et al (2015) have obtained another confirmation of the depth of the meridional flow's shallow reversal. Because the above three measurements are independent and give the same depth, we judge that they together constitute solid evidence that the meridional flow reverses at about 0.92 $R_{Sun}$.

The helioseismic time-distance analysis of GONG Doppler data by Jackiewicz et al (2015) and a helioseismic time-distance analysis of HMI Doppler data by Rajaguru & Antia (2015) each cast doubt on the veracity of both the 10 m s$^{-1}$ speed found by Zhao et al (2013) for the shallow return meridional flow and the deeper second reversal found in the meridional flow by Zhao et al (2013).

*2.3. Recently-Found Evidence Supporting Three of Babcock's Assumptions*

From nine consecutive sunspot cycles (Cycles 15-23) in each polar hemisphere, via observations of the magnetic flux in the polar cap, the sunspot area, and the tilt of the sunspot active regions, Munoz-Jaramillo et al (2012, 2013) obtained the following three results. First, in each polar hemisphere the sunspot-area amplitude of a sunspot cycle is strongly correlated with the polar-cap magnetic flux above 70° latitude at the sunspot-area minimum at the beginning of that sunspot cycle. Second, in each polar hemisphere the sunspot-area amplitude of a cycle is not correlated with that cycle's average polar flux above 70° latitude, nor with the average polar flux of two cycles ago, nor with the average polar flux of three cycles ago. Third, in each polar hemisphere the polar flux above 70° latitude at the sunspot-area minimum at the end of a sunspot cycle is strongly correlated with the product of the sunspot-area-weighted average tilt of the sunspot regions of that cycle, but is not correlated with that cycle's sunspot-area amplitude alone. As is tacitly recognized by Munoz-Jaramillo et al (2013), these observational results are evidence supporting the following three assumptions of Babcock's dynamo scenario. First, the assumption that in each polar hemisphere the toroidal source field of each sunspot cycle is built from the polar field of the preceding cycle by shearing of the polar field's horizontal component in the convection zone by the latitudinal differential rotation of the convection zone. Second, the assumption that in each polar hemisphere the unipolar field that cancels the polar field of the preceding sunspot cycle during the rise and maximum of a sunspot cycle (and occupies the polar cap after the cycle's maximum) is the net active-region following-polarity field left in that hemisphere by cancelation of active-region leading-polarity field across the equator with opposite-polarity leading-polarity field from the opposite hemisphere. Third, in each polar hemisphere the remnant net following-polarity field is swept into the polar cap from lower latitudes by poleward meridional flow of the top of the convection zone.

3. ASSUMPTIONS

Three assumptions on which our solar dynamo scenario is based are the three assumptions of Babcock stated above, in the last three sentences of Section 2. For our scenario we also employ the following additional assumptions.

The following three assumptions we take from Spruit & Roberts (1983) and Spruit (2010). First, we assume that the band of toroidal field from which the emerged Ω loops in a sunspot band arise resides at the bottom of the convection zone and is built up there by shearing of horizontal poloidal field by the latitudinal differential rotation of the bottom of the convection zone. [Although recent simulations by Nelson et al (2013, 2014) of dynamo action in the interior of the turbulent convection zone indicate that it might be possible for the toroidal source field of the active-region Ω loops to be generated and kept in the interior of the convection zone, an essential premise of our scenario is that the toroidal source field is generated and kept at the bottom of the convection zone. Because toroidal magnetic flux tubes are buoyantly unstable in the Sun's convection zone, it is reasonable to expect that the bands of internal toroidal field from which the large-scale Ω loops presumably arise do not sit in the middle of the convection zone but at the bottom, that is, at the interface between the convection zone and the radiative interior (Spruit & van Ballegooijen 1982). Moreover, MHD simulations of the rise through the convection zone of large-scale Ω loops from buoyantly unstable flux ropes at the bottom of the convection zone indicate that Joy's law results from the Coriolis force acting on the rising Ω loops (Fan et al 1994). Analytical modeling of the buffeting of the rising Ω loops by the turbulent convection indicates that the observed scatter in the tilt about the average results from the buffeting (Loncope & Choudhuri 2002). The agreement of these simulation and modeling results with the observed tilts strengthens the premise of our scenario that the Sun's large-scale Ω loops come from toroidal field generated at the base of the convection zone by the global dynamo process.] Second, we assume that, as it disperses, the field of an emerged Ω loop remains connected to the toroidal field band from which it arose. Third, we assume that the convection zone's free convection, by its buoyancy forces and downward-pushing topological pumping acting on the magnetic field's horizontal component, expels from the interior of the convection zone the horizontal field component of each connecting flux tube quickly enough to keep the connecting flux tubes roughly vertical in the interior of the convection zone and nearly horizontal along the bottom of the convection zone. ["Free convection" is the convection that results from buoyancy instability. Topological pumping is also called turbulent pumping (Charbonneau 2010). A "connecting flux tube" connects a unipolar clump of photospheric magnetic flux to the toroidal field band from which the field in the flux tube arose.]

We also make the following trio of assumptions concerning reconnection and cancellation of oppositely-directed merging magnetic fields in the photosphere and convection zone. We assume that when opposite-polarity photospheric flux clumps connected to the same toroidal field band meet and cancel, the resulting subsurface flux loop retracts back and melds with that toroidal field band. Similarly, we assume that when a photospheric flux clump connected to one toroidal field band meets and cancels with an opposite-polarity flux clump connected to another toroidal field band, the resulting flux loop connecting the two toroidal field bands is forced to the bottom of the convection zone and kept there by the topological pumping of the free convection a la Spruit & Roberts (1983). The third related assumption is that when two oppositely-directed horizontal fields meet at or near the bottom of the convection zone, they cancel via reconnection.

Finally, we assume that at the bottom of the convection zone in each polar hemisphere there is equatorward meridional flow that carries each toroidal field band equatorward at the observed drift rate of the corresponding torsional-oscillation wave and sunspot band ($\sim 1$ m s$^{-1}$). For this assumption to be viable, we are compelled to assume that instead of the meridional flow being composed of only a single cell that spans the hemisphere and fills the convection zone from bottom to top, it is composed of a stack of three cells. As discussed in Section 2, we take as observationally established that with increasing depth from the top the meridional flow reverses direction at a depth of about 0.08 $R_{Sun}$. From consideration of mass-flow balance, because the depth-averaged speed of the poleward meridional flow above the reversal is less than 10 m s$^{-1}$, and because the depth-averaged mass density of the convection zone is >100 times larger from the reversal to the bottom than from the reversal to the top, the entire convection zone below the reversal cannot be flowing equatorward as fast as ~ 1 m s$^{-1}$. So, in order to have the meridional flow at the bottom of the convection zone be equatorward at a speed of ~ 1 m s$^{-1}$, we assume there are three stacked cells of meridional flow. If there are three stacked cells, then the time-distance helioseismic results of Zhao et al (2013) tentively indicate that the three cells each have roughly the same thickness, ~ 0.1 $R_{Sun}$. In depicting our scenario, for simplicity of illustration, in each polar hemisphere we draw three cells of equal thickness filling the 0.3 $R_{Sun}$ thickness of the convection zone, as in Figure 1.

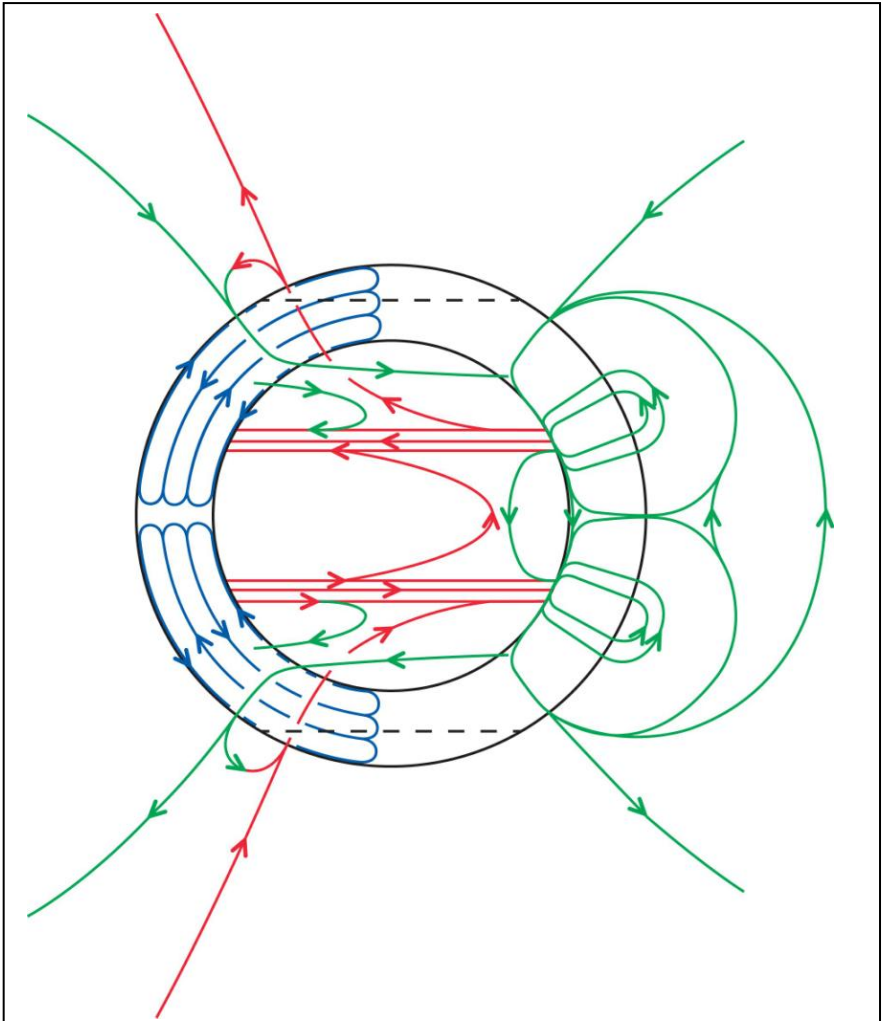

**Figure 1.** Schematic depiction of the proposed three-dimensional configuration of the Sun's global magnetic field in 1998, in the rise of sunspot Cycle 23. Blue lines represent meridional flow in the convection zone. Red lines represent the Cycle-22 polar-cap field and its solar-interior extent from which the two red toroidal source-field bands for the sunspot-active-region Ω loops of Cycle 23 have been built at the bottom of the convection zone by latitudinal differential rotation. The green lines represent field from emerged sunspot-active-region Ω loops – which arch predominantly east-west but tilt equatorward – that have dispersed enough that some of their leading-polarity flux from opposite polar hemispheres has cancelled by reconnection at the equator, and some of the resulting net following-polarity flux in each polar hemisphere has been swept poleward by the meridional flow of the top of the convection zone to encounter the red polar cap field. The green closed lines above the west limb represent the above-surface plane-of-the-sky projection of dispersing sunspot-active-region Ω loops that are centered on the west limb and are viewed nearly end-on from their following-polarity ends.

While the convection zone in each polar hemisphere possibly is filled by a stack of three cells of meridional flow of about equal thickness, the speed of the equatorward flow in the lower part of the top cell and the upper part of the middle cell remains uncertain and the speed of the poleward flow in the lower part of the middle cell and the upper part of the bottom cell remains unknown (see Section 2). Even so, it is reasonable to expect each of these speeds to be slower than the ~ 10 m s$^{-1}$ poleward speed of the top of the convection zone and faster than the ~ 1 m s$^{-1}$ speed of our assumed eqatorward meridional flow at the bottom of the convection zone. The three reversals of meridional flow continually act to deform the above-defined connecting flux tubes from being nearly vertical in the convection zone. Their deformation of a flux tube from vertical is in the meridional plane through the flux tube and (viewed from the west in the northern hemisphere) has the shape and progression of an increasingly more extreme S curve having its top and bottom at the top and bottom of the convection zone. In addition to these three anticipated meridional horizontal shear flows, the convection zone has a certified near-surface horizontal shear-flow layer in its rotation (Howe 2009). At all latitudes up to at least 70°, the rotation speed increases with depth below the surface down to a depth of about 50,000 km. The rotation-speed difference between this depth and the surface is about 40 m s$^{-1}$ at the equator and steadily decreases with latitude to about 10 m s$^{-1}$ at 70° (Hathaway 2012a). The shear flow in this layer continually acts to shear the connecting flux tubes from vertical, acts to give them an increasingly greater east-west (toroidal) horizontal component inside the layer.

We assume that the free convection in the convection zone, via its buoyancy forces and topological pumping, continually acts to remove the cross-radial deformations given to the connecting flux tubes by the meridional shear flows and by the near-surface-rotation shear flow, and acts enough faster than these shear flows that it keeps the connecting flux tubes roughly vertical in the convection zone. How realistic this assumption is in the Sun's convection zone is an open question for investigation by equations-based magneto-convection simulations beyond the scope of this paper.

The main new feature of our amendment of Babcock's dynamo scenario has two aspects. One aspect is that as the net following-polarity magnetic field of a dispersed emerged active-region Ω loop is swept to polar latitudes by the meridional flow of the top of the convection zone, it stays connected to its toroidal source-field band at the bottom of the convection zone. The other aspect is that buoyancy and topological pumping by the free convection in the convection zone force the poleward-migrating net following-polarity field to take the following path in threading through the convection zone to its source-field band: from the surface, the poleward-migrating field runs roughly vertically down to near the bottom of the convection zone and then turns 90° and runs nearly horizontally westward and equatorward along the bottom, winding back to its source-field band. So far as we are aware, our scenario is new in that there is no previous scenario or MHD simulation of a flux-transport solar dynamo that has both of these aspects of our scenario.

## 4. SCHEMATIC OF THE MAGNETIC CYCLE

In this section we lay out our schematic of the Sun's global magnetic cycle with four successive cartoons, outlining the proposed three-dimensional configuration and evolution of the global magnetic field above the convection zone and in the interior and bottom of the convection zone, in and between four successive phases. The first three phases are the rise phase, maximum phase, and decline phase of one sunspot cycle, and the fourth phase is the rise phase of the next sunspot cycle. In each of the first three cartoons, the field direction, shown by arrowheads on the field lines, matches the direction of the global field observed in and above the photosphere surface in the corresponding phase of sunspot Cycle 23, the Cycle that began in 1996 and ended in 2008 (e.g., Hathaway 2010). In the fourth cartoon, the field direction matches that of the observed global field in the rise of the next sunspot cycle, Cycle 24.

### *4.1. Phase I*

The cartoon in Figure 1 depicts our proposed 3-D global magnetic field configuration in 1998, in the rise of sunspot Cycle 23, right after the cessation of the two sunspot bands of Cycle 22 as they met at the equator. The view direction of the cartoon is orthogonal to the Sun's rotation axis. Solar north is up and west is to the right. The cartoon shows both (1) the cross section of the convection zone in the plane that is orthogonal to the view direction and passes through the center of the Sun and (2) the projection, on that plane, of the facing hemisphere of the bottom of the convection zone. Magnetic field lines are shown in projection on that plane. Each dashed line parallel to the equator is the projection of a polar cap's edge at latitude 60°, north or south. The assumed three stacked cells of meridional flow in the north and south hemispheres of the convection zone are shown by the blue stream lines with the flow direction shown by arrow heads. Lines of polar magnetic field from which the Cycle-23 toroidal field bands have been built at the bottom of the convection zone by latitudinal differential rotation are red. Lines of magnetic field from emerged Cycle-23 large-scale Ω loops are green.

At this time in Cycle 23, the field had positive polarity in the northern polar cap and negative polarity in the southern polar cap. Only one representative line of each polar-cap field is drawn in this cartoon. At the time in Cycle 23 depicted in Figure 1, Ω loops that made sunspots had been emerging from the red toroidal field band in each hemisphere for about two years. We suppose that by about a year before the time depicted in Figure 1, due to the on-average equatorward tilt of these predominantly east-west emerged Ω loops, as the Ω-loop field dispersed, net leading-polarity magnetic flux began canceling across the equator. By the time of Figure 1, resulting net following-polarity flux in each hemisphere had been swept to the polar cap by the meridional flow of the top of the convection zone, and began canceling the opposite-polarity flux that occupied the polar caps in this phase of Cycle 23. In Figure 1, the cancelation of the old polar-cap field (red) by the encroaching new following-polarity field (green) is symbolized by the half-red/half-green field-line arch across the east edge of each polar cap.

As is illustrated in Figure 1, we suppose, following Spruit & Roberts (1983), that (1) as emerged Ω-loop following-polarity field was carried to a polar cap, it remained connected to the toroidal field band from which the Ω loops emerged, and (2) topological pumping by the free convection in the convection zone kept this field nearly vertical above the bottom of the convection zone and nearly horizontal along the bottom, where this field winds back to its source toroidal field band. By the time of Figure 1, the horizontal stretch of the poleward-migrating following-polarity emerged field was sheared enough by the latitudinal differential rotation in the bottom of the convection zone that the new oppositely-directed toroidal field band (green) was building up near the 60° edge of the polar cap, i.e., the new toroidal band was centered at about 55° latitude in each hemisphere, as is depicted in Figure 1. [From 30° to 60° latitude, the meridional flow of the top of the convection zone is roughly constant at about 10 m s$^{-1}$, and the angular rotation rate of the top of the convection zone decreases roughly linearly with increasing latitude, from 13.88 deg day$^{-1}$ at 30° to 11.65 deg day$^{-1}$ at 60° (Hathaway & Rightmire 2011). From these observations, we obtain (1) 1.1 years for the meridional-flow transit time from 30° to 60°, and (2) 1.1 turns per year for the average rate of longitudinal winding, of a field line connecting an element of following-polarity flux to the toroidal source-field band near 30° latitude, during the flux element's transit from 30° to 60°. We thus estimate that the connecting field line acquired 1.2 turns of longitudinal winding during the 30°-60° transit of the following-polarity flux element. Accordingly, in Figure 1 we have given about 1.2 turns of longitudinal winding to the representative green field line that connects the following-polarity flux near 60° latitude to the red toroidal source-field band near 30° latitude.] In Figure 1, the new high-latitude toroidal field band in each hemisphere is represented by the single green field line that runs east-west across the bottom of the convection zone. This new toroidal field band (green) and the old toroidal field band (red) are both carried equatorward by the meridional flow at the bottom of the convection zone.

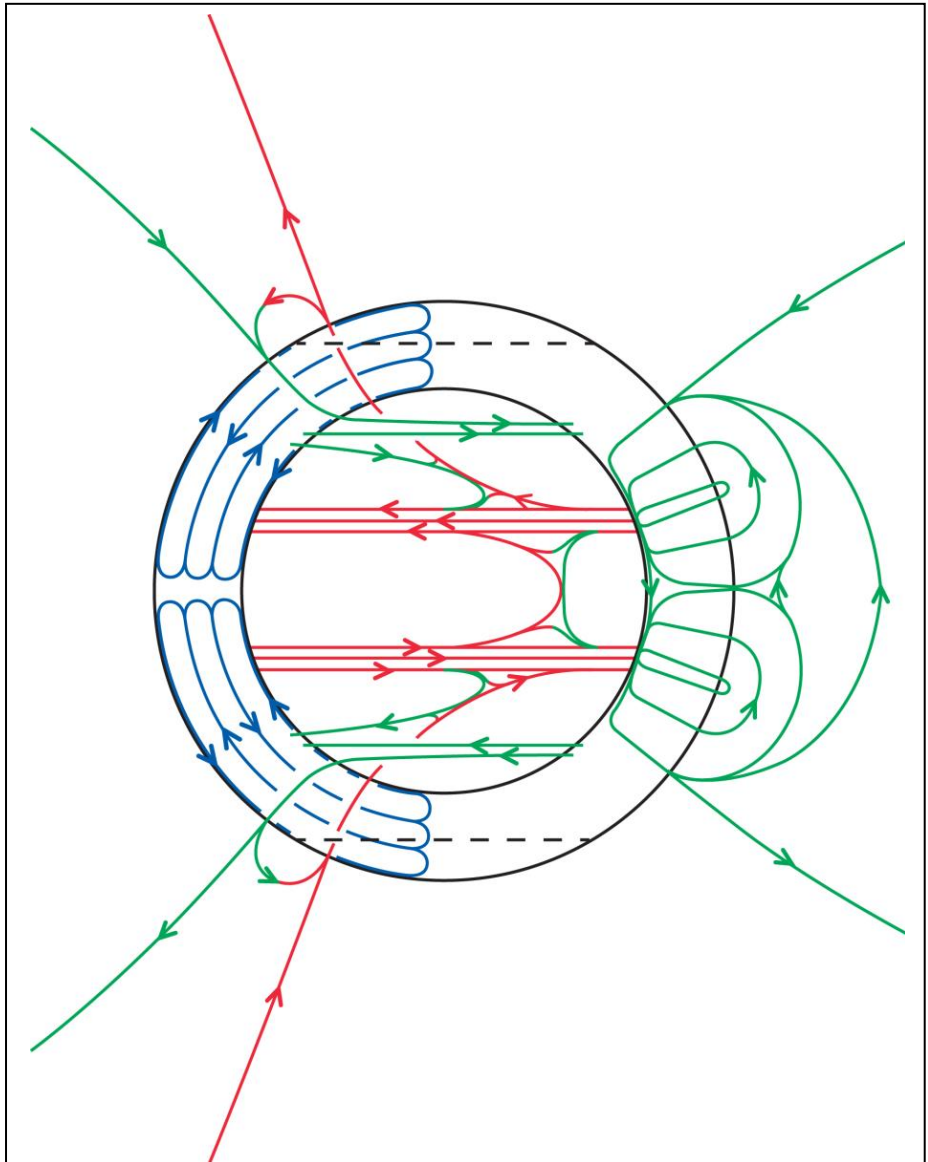

**Figure 2.** Proposed three-dimensional global field configuration in 1999, late in the rise of sunspot Cycle 23. By this time in our scenario, both poleward and equatorward of each (red) toroidal source-field band for the sunspot-active-region Ω loops of the rising sunspot cycle, topological pumping by the free convection in the convection zone has pushed the oldest of the new (green) horizontal field down against the old (red) horizontal field below, forcing these two oppositely-directed horizontal fields to cancel via reconnection such as that symbolically depicted here.

As is depicted in Figure 1, flux-cancelation reconnection of two opposite leading-polarity field lines at the equator produces a new (reconnected) field line above and a corresponding new field line below. The upward new field line rises into the corona and solar wind and is rooted at opposite ends in poleward-migrating following-polarity flux in opposite hemispheres, north and south. The downward new field line connects the two Cycle-23 toroidal field bands across the equator and is pushed to the bottom of the convection zone by topological pumping by the free convection a la Spruit & Roberts (1983). The polar (north-south) direction of this new (green) connecting field is opposite that of the old (red) cross-equator connecting field from which the equatorward halves of the two Cycle-23 (red) toroidal field bands were built by latitudinal differential rotation. Similarly, the polar direction of the new (green) field at the bottom of the convection zone on the poleward side of each Cycle-23 (red) toroidal field band is opposite that of the old (red) polar field from which the poleward halves of the two Cycle-23 toroidal field bands were built.

In Figure 1, we suppose that for a year or so the new (green) horizontal field at the bottom of the convection zone, both equatorward and poleward of each Cycle-23 toroidal field band, lies a little higher above the bottom of the convection zone than the old (red) horizontal field at these latitudes. During that time, we suppose the new horizontal field is gradually pushed farther down by topological pumping by the free convection to finally meet the old horizontal field and thus cancel the poloidal component of the old horizontal field. In our scenario, at the time in sunspot Cycle 23 depicted in Figure 1, during the steep rise to that Cycle's maximum sunspot number, little if any of this cancelation has yet occurred, and hence each old (red) toroidal field band is still nearly entirely connected on its poleward side by old (red) horizontal field to the old (red) polar-cap field, and is still nearly entirely connected across the equator by old (red) horizontal field to the other old toroidal field band. So, the cartoon in Figure 1 is meant to portray that in the steep rise of a sunspot cycle to its maximum phase, the two toroidal field bands from which the sunspot-making $\Omega$ loops are emerging are continuing to be built up by the shearing of the poloidal component of the old horizontal field at the bottom of the convection zone by the latitudinal differential rotation of the bottom of the convection zone.

*4.2. Phase II*

Figure 2 shows the proposed global field configuration in 1999, about a year later in sunspot Cycle 23 than the time depicted in Figure 1. In 1999, the two sunspot bands were each about 20° in latitude from the equator, north and south, and sunspot Cycle 23 was entering its 1999-2002 maximum phase (Hathaway 2010; Hathaway & Rightmire 2011). Figure 2 depicts that during the roughly one-year increment of Cycle 23 from Figure 1 to Figure 2, the meridional flow in each polar hemisphere of the bottom of the convection zone gradually carried both of the two toroidal field bands (the old red one and the new green one) closer to the equator by a few degrees in latitude. During this time, equatorward-tilted sunspot-making Ω loops kept emerging from the red toroidal field band in each hemisphere and dispersed, resulting in net cancelation of leading-polarity flux across the equator, leaving in each hemisphere net following-polarity flux that was continually swept to the polar cap by the surface meridional flow and continually canceled and replaced the opposite-polarity old polar flux. In this manner, by 2002 (three years after the time depicted in Figure 2) in each hemisphere the old polar field in the polar cap had been completely canceled and replaced by the opposite-polarity poleward-migrating following-polarity flux of Cycle 23 in that hemisphere (Hathaway 2010).

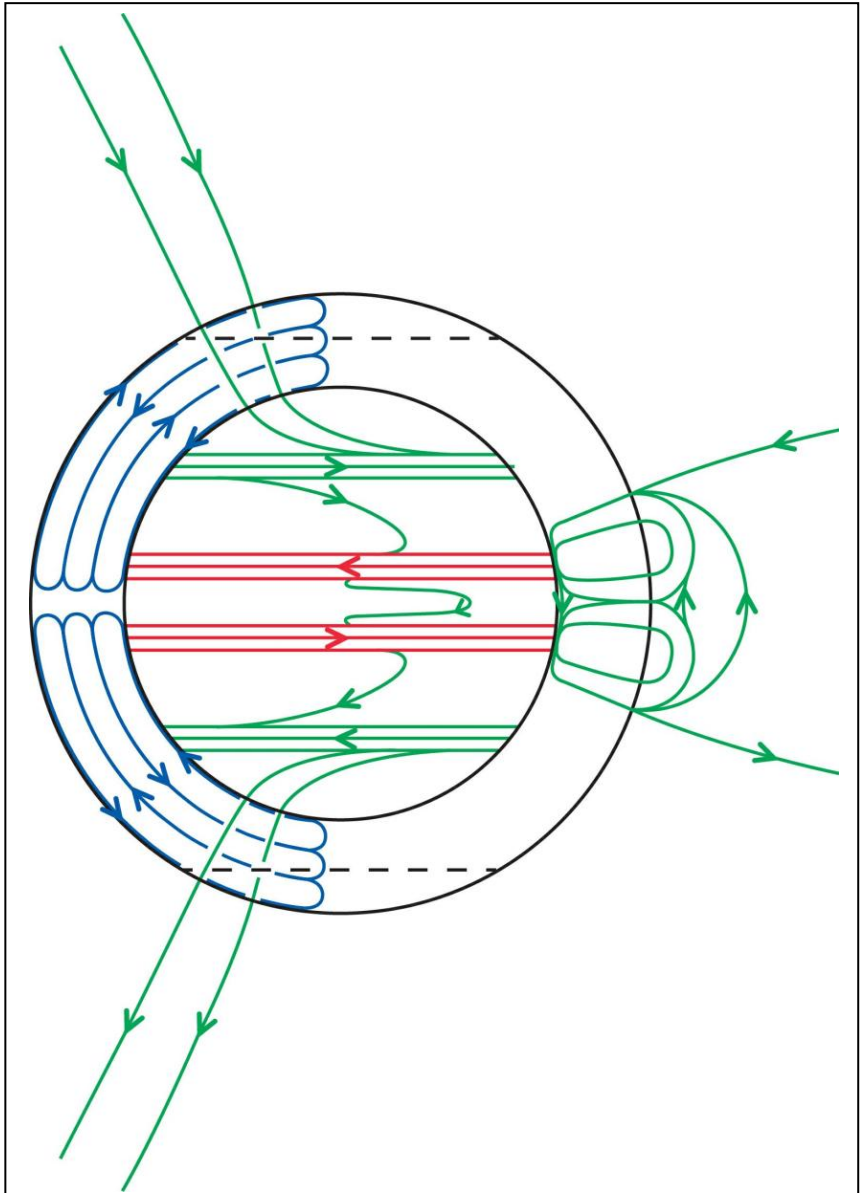

**Figure 3.** Proposed three-dimensional global field configuration in 2005, midway in the decline of sunspot Cycle 23. At this time in our scenario, the old (red) toroidal source-field bands of the ongoing sunspot cycle are decaying, being eroded on each side by oppositely-directed new (green) fields connected to them. The new (green) toroidal source-field bands for the coming sunspot cycle are continuing to grow and are the source of ephemeral-active-region Ω loops, but have not yet grown enough for sunspot-active-region Ω loops to arise from them.

Figure 2 also depicts that as sunspot Cycle 23 entered its maximum phase the latitudinal differential rotation of the bottom of the convection zone continued to build up the new (green) toroidal field band in each polar hemisphere. As is also depicted in Figure 2, in 1999 in our scenario there was in progress, via reconnection at the bottom of the convection zone, appreciable cancelation of the poloidal component of the old (red) horizontal field stemming from each old (red) toroidal field band by the poloidal component of the new (green) horizontal field steming from that old toroidal field band. That is, on the poleward side of each old toroidal field band, by 1999 an appreciable amount of old horizontal field connecting the old polar-cap field to the old toroidal field band had been disconnected from the old toroidal field band by reconnection with new horizontal field connecting the new toroidal field band to the old toroidal field band, as depicted in Figure 2. Simultaneously in our scenario, about the same amount of old (red) horizontal field connecting the two old toroidal field bands across the equator had been disconnected by reconnection (as depicted in Figure 2) with new (green) horizontal field that also had connected the two old toroidal field bands across the equator. In our scenario, from this reconnection process at the bottom of the convection zone, by 2002, after the old (red) polar field in each polar cap had been completely canceled and replaced by new (green) opposite-polarity polar field, the poloidal component of all of the old (red) horizontal field stemming from the sides of the old toroidal field bands had been canceled by the poloidal component of the new horizontal field stemming from the sides of the old toroidal field bands. Consequently, from then on until 2008 in the late declining phase of Cycle 23, each toroidal field band from which the sunspot $\Omega$ loops of Cycle 23 emerged was, on its poleward side, entirely connected by new horizontal field to the new toroidal field band, and was, on its equatorward side, entirely connected across the equator to the other old toroidal field band by new horizontal field (as in Figure 3). These considerations imply that by 1999 the building-up of the two old toriodal field bands by the latitudinal differential rotation of the base of the convection zone was waning (Figure 2), by 2002 the building-up of the old toroidal field bands had stopped, and thereafter in the declining phase of Cycle 23 (Figure 3) the field in the old toroidal bands diminished and finally canceled when the two bands met at the equator (resulting in the global field configuration shown in Figure 4).

*4.3. Phase III*

Figure 3 shows the proposed global field configuration in 2005. At that time sunspot Cycle 23 was midway through its 2002-2008 declining phase and its two equatorward-migrating sunspot bands were each centered at about 10° in latitude from the equator, as is depicted in Figure 3 (Hathaway 2010; Hathaway & Rightmire 2011). In the manner depicted in Figure 3, throughout the decline of Cycle 23 the polarity of each polar-cap field was that of the trailing feet of the Cycle-23 large-scale $\Omega$ loops arising from the red toroidal field bands. Throughout this interval, all of the large-scale field in each polar cap was connected to its hemisphere's new (green) toroidal field band (which by 2005 had descended in latitude to be centered at about 40° in latitude from the equator, as in Figure 3). Also throughout this time, each old (red) toroidal field band was connected poleward only to the new toroidal field and only by new (green) horizontal field, and the two old toroidal field bands were connected across the equator only by new (green) horizontal field (as in Figure 3). In our scenario, throughout the declining phase of Cycle 23, due to the field linkages shown in Figure 3, the latitudinal differential rotation of the base of the convection zone in each polar hemisphere continued to build up the field in the new (green) toroidal band but not in the old (red) toroidal band.

As was noted in Section 2, throughout the maximum and declining phase of each sunspot cycle, the extended-solar-cycle precursor of the next sunspot cycle is in progress poleward of the sunspot bands of the ongoing sunspot cycle (Wilson et al 1988; McItosh et al 2014). In each polar hemisphere, starting early in the maximum phase of the ongoing sunspot cycle, when the center latitude of the sunspot band is about 25°, and continuing through the declining phase of the sunspot cycle, there is an equatorward-drifting broad band of sunspotless ephemeral active regions centered about 30° poleward of the sunspot band. The predominant east-west polarity direction of the ephemeral active regions in this high-latitude band is that of the sunspot active regions of the coming sunspot cycle. Those sunspot active regions begin arising in the ephemeral-region band when that band has descended to about 30° latitude. This observed progression of the extended solar cycle indicates that the toroidal source-field band for each sunspot band is born at a latitude of about 55° early in the maximum phase of the previous sunspot cycle, and suggests that source-field band grows in field strength and flux as it drifts equatorward. The onset of the production of sunspot active regions when the ephemeral-region band has descended to 30° suggests that throughout its drift from 55° to 30° the source-field band is big enough and strong enough for ephemeral-active-region $\Omega$ loops to arise from it, but has not yet grown enough for sunspot-active-region $\Omega$ loops to arise from it, and that by the time the source-field band has descended to 30° and below it has grown enough for both ephemeral-active-region $\Omega$ loops and sunspot-region $\Omega$ loops to arise from it (Wilson et al 1988; McIntosh et al 2014). In our scenario, in each polar hemisphere the generation and evolution of the toroidal source-field band of the coming sunspot cycle (the green toroidal field band in Figures 1, 2, 3, and 4) fits this interpretation of the observed extended solar activity cycle. That is, in our scenario the new source-field band is born centered at about 55° latitude in the rise of the ongoing sunspot cycle (Figure 1) and as it drifts equatorward it is continually amplified by the latitudinal differential rotation (Figures 2, 3, and 4) until the reversal of the polar-cap field in the maximum of the coming sunspot cycle. In that qualitative way, our scenario fits the observed progression of the extended solar activity cycle. Our scenario does not specify why sunspot-region $\Omega$ loops start emerging from a new-cycle source-field band when the band has descended in latitude to about 30° instead of starting earlier or later in the growing source-field band's descent in latitude. That question can be addressed only by MHD simulations of our scenario beyond the scope of this paper.

Also throughout the declining phase of each sunspot cycle in our scenario, as Figure 3 indicates, the differential rotation acting on the new horizontal field connecting the two old toroidal bands across the equator produces in each hemisphere new toroidal field of the opposite direction to that of the old toroidal field in that hemisphere. That new toroidal field presumably erodes the equatorward side of each old toroidal field band via reconnection. Likewise, as Figure 3 shows, the direction of the new toroidal field being generated on the poleward side of each old toroidal field band is opposite that of the old band and so possibly erodes the poleward side of each old band via reconnection.

*4.4. Phase IV*

Figure 4 shows the proposed global field configuration in 2009. The last sunspots of Cycle 23 occurred in mid 2008 as the two sunspot bands of Cycle 23 merged at the equator. The first sunspots of Cycle 24 occurred early in 2008 when the two sunspot bands of Cycle 24 were each centered at about 30° in latitude from the equator (Hathaway & Rightmire 2011). By mid 2009, each of these two bands of sunspots had migrated equatorward by about 5° to be centered at about 25° from the equator, each centered radially above the corresponding (green) toroidal field band drawn in Figure 4. These two field bands are the toroidal field bands from which sunspot-making $\Omega$ loops of Cycle 24 were arising.

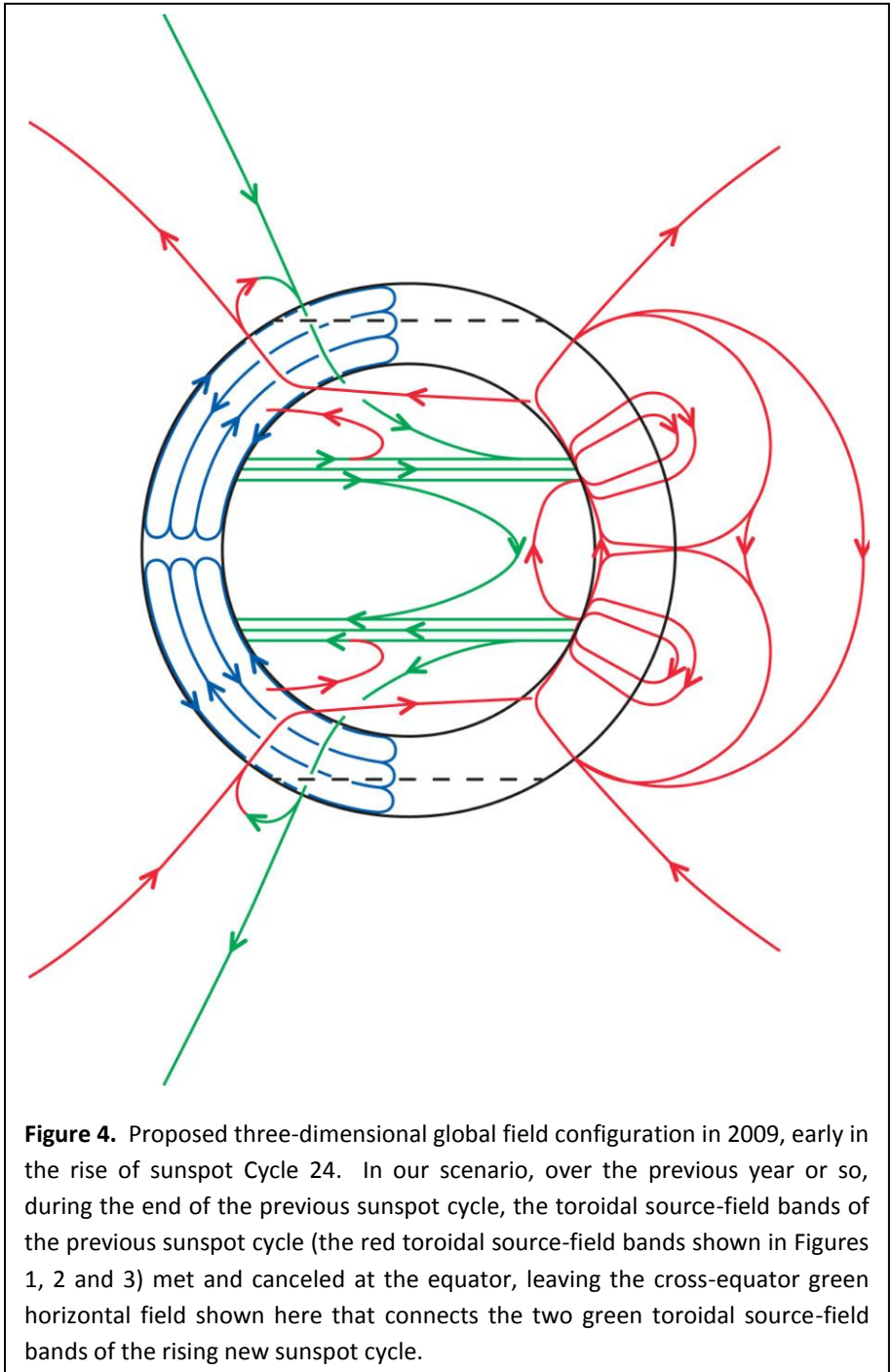

**Figure 4.** Proposed three-dimensional global field configuration in 2009, early in the rise of sunspot Cycle 24. In our scenario, over the previous year or so, during the end of the previous sunspot cycle, the toroidal source-field bands of the previous sunspot cycle (the red toroidal source-field bands shown in Figures 1, 2 and 3) met and canceled at the equator, leaving the cross-equator green horizontal field shown here that connects the two green toroidal source-field bands of the rising new sunspot cycle.

The cartoon in Figure 4 is a repeat of the cartoon in Figure 1, except that the red field lines in Figure 1 are the green ones in Figure 4 and the green field lines in Figure 1 are the red ones in Figure 4. In our scenario, by the time in Cycle 24 depicted in Figure 4, the two red toroidal field bands of Cycle 23 had completely canceled at the equator and this cancelation left the two green Cycle-24 toroidal field bands connected across the equator predominantly by green horizontal field that was at the bottom of the convection zone during Cycle 23 and stemmed from the poleward side of each Cycle-23 (red) toroidal field band, as depicted in Figure 3. Figure 4 shows that in 2009 in our scenario the two Cycle-24 toroidal field bands had some connection across the equator at the bottom of the convection zone by new red horizontal field that had come from the topological-pumping-driven sinking of Cycle-24 subsurface transequatorial field loops made by cancelation reconnection of Cycle-24 opposite leading-polarity flux at the equator. Figure 4 is also intended to indicate that on its poleward side each green toroidal field band is predominantly connected to the old (green) polar-cap field, but has some connection to the new (red) poleward-migrating net following-polarity flux from the dispersed fields of the large-scale Ω loops of Cycle 24.  Figure 4 also depicts that by 2009, some Cycle-24 net following-polarity flux in each hemisphere had reached the polar cap and was canceling the old (green) Cycle 23 polar-cap field, and latitudinal differential rotation of the bottom of the convection zone had begun building the (red) toroidal field band for sunspot Cycle 25.  In each polar hemisphere in Figure 4, that band is centered about 55° in latitude from the equator.

In the manner depicted in Figures 1-4, our proposed global magnetic field configuration evolves over each sunspot cycle to become the global field configuration of the next sunspot cycle.

5. SUMMARY AND DISCUSSION

Building on Babcock (1961), by using insights of Spruit and his collaborators for the convective expulsion of the horizontal component of large-scale magnetic field from the convection zone (Spruit & van Ballegooijen 1982; Spruit & Roberts 1983; Spruit 2011), and based on recent observations of the differential rotation of the convection zone (Howe 2009), stacked cells of meridional flow in the convection zone (Hathaway 2012; Zhao et al 2013), the recent confirmation of the strong correlation of the amplitude of a sunspot cycle with the amount of polar-cap field present at the start of that sunspot cycle (Munoz-Jaramillo et al 2013), and the recent confirmation of the extended solar activity cycle [the occurrence in each polar hemisphere of the broad latitudinal band of magnetic activity that arises around 55° latitude and migrates equatorward over the course of each sunspot cycle to become the sunspot band of the next sunspot cycle (McIntosh et al 2014)], we have devised a schematic visualization of the Sun's global dynamo process.  In this scenario, the toroidal field bands that give rise to the sunspot bands of each sunspot cycle are built at the bottom of the convection zone from the polar field of the preceding sunspot cycle by the latitudinal differential rotation of the bottom of the convection zone.  In each polar hemisphere, the toroidal field band that becomes the source field of the next-cycle sunspot band is initially centered at about 55° in latitude from the equator and begins building up there at the bottom of the convection zone as soon as some of that hemisphere's net following-polarity field from the sunspot-making Ω loops of the ongoing sunspot cycle has been swept to the edge of the polar cap by the surface meridional flow, a year or so after the first of those Ω loops have emerged.    Emergence of ephemeral-active-region Ω loops from the equatorward-migrating next-sunspot-cycle growing toroidal source-field band over the 8-9 year remainder of the ongoing sunspot cycle manifests the high-latitude interval of the extended solar activity cycle, during which the latitudinal center of the growing toroidal field band descends from 55° to 30° in latitude from the equator, prior to the emergence of the first sunspot-making Ω loops from the growing toroidal source-field band.

In our scenario, in each polar hemisphere, the growing next-sunspot-cycle toroidal source-field band is at the bottom of the convection zone poleward of the toroidal source-field band of the ongoing sunspot cycle because it is made there by the latitudinal differential rotation's shearing of the horizontal interval of the "high-latitude" net following-polarity field that has migrated to and into the polar cap while remaining connected to the source-field band of the ongoing sunspot cycle.  The horizontal interval of that connecting field is kept pushed to the bottom of the convection zone by the free convection's topological pumping. That results, as in Figure 3, in the growing next-sunspot-cycle toroidal source-field band being: (a) connected on its higher-latitude side to that side's horizontal interval of the high-latitude net following-polarity field, and (b) connected on its lower-latitude side to the toroidal source-field band of the ongoing sunspot cycle by that side's horizontal interval of the high-latitude net following-polarity field.

Our scenario amends Babcock's original scenario for the global dynamo process engendering the Sun's 22-year magnetic cycle (Babcock 1961). The main amendment is that the Sun's global internal poloidal field does not run directly from polar cap to polar cap at some time a few years prior to the onset of the next sunspot cycle, as Babcock assumed. Instead, in our scenario, there are always two oppositely-directed equatorward-migrating toroidal field bands in each polar hemisphere throughout each sunspot cycle: one is the band from which the sunspot-making $\Omega$ loops of the present sunspot cycle are arising, and the other is the band from which the sunspot-making $\Omega$ loops of the next sunspot cycle will begin arising a year or so before the end of the present sunspot cycle. Our scenario retains Babcock's assumption that the two toroidal field bands from which arise the sunspots of each sunspot cycle are built from global poloidal field by latitudinal differential rotation inside the Sun. The corresponding amendment in our scenario is that the toroidal field bands buildup at the bottom of the convection zone rather than in the interior of the convection zone as Babcock supposed.

From their inspection of the extended solar activity cycle, McIntosh et al (2014) and McIntosh & Leamon (2014) propose that the two oppositely-directed toroidal field bands in each polar hemisphere destructively interact with each other, and that the lower-latitude toroidal field band in each polar hemisphere destructively interacts across the equator with the oppositely-directed lower-latitude toroidal field band in the other hemisphere progressively more as these two bands approach each other and finally merge and cancel at the equator. This idea in turn suggests that each sunspot-producing toroidal field band grows most rapidly during the rise of each sunspot cycle because the high-latitude toroidal field band in each polar hemisphere is still weak (is still in its early growth phase) and because the two sunspot-producing toroidal field bands are still far apart across the equator, as in Figures 1 and 2. Our global dynamo scenario is consistent with this suggestion and visualizes a prospective mechanism for the "destructive interaction" between adjacent oppositely-directed toroidal field bands. In our scenario, in the rise of a sunspot cycle, as in Figures 1 and 2, the sunspot-producing toroidal field bands are each still mainly connected poleward to the old polar-cap field from the previous sunspot cycle and are each still mainly connected across the equator to the other sunspot-producing toroidal field band by old horizontal field from the previous solar cycle, and are therefore growing via latitudinal differential rotation of the old horizontal poloidal field still connected to them. In the decline of a sunspot cycle, as in Figure 3, the two sunspot-producing toroidal field bands are each connected poleward entirely to the new high-latitude toroidal field band by new horizontal field from the ongoing sunspot cycle and are each connected across the equator to the other sunspot-producing toroidal field band entirely by new horizontal field from the ongoing sunspot cycle. Consequently, in the decline of a sunspot cycle in our scenario the sunspot-producing toroidal field bands are no longer growing and are being eroded by oppositely-directed toroidal field being generated on their poleward sides and on their equatorward sides by the latitudinal differential rotation of the bottom of the convection zone, as is depicted in Figure 3.

X-ray jets in the polar coronal holes are evidently made by a burst of reconnection of closed magnetic field in a jet's base with the ambient open field of the coronal hole (e.g., Cirtain et al 2007).  It is observed that during the eruption of most of these jets, a bright point is produced at one edge of the base and the jet spire drifts away from the bright point (Savcheva et al 2009).  If the closed field in the base of the jet is that of a closed magnetic bipole (bipolar ephemeral active region) that has emerged in the coronal hole, Sterling et al (2015) have recently discovered that the base-edge bright point sits on the outside edge of the minority-polarity end of the bipole (the end having the polarity opposite to that of the ambient open field), and is produced by internal reconnection of the legs of the erupting magnetic field in the eruption of a small filament from the site of the bright point.  The production of the jet's spire by external reconnection of the filament-carrying erupting field with the ambient open field explains why the spire drifts away from the bright point (Sterling et al 2015).  Previously, Savcheva et al (2009), by assuming that the base of the coronal-hole X-ray jet is an emerging bipole, that the bright point occurs on the minority-polarity end of the bipole, and that the spire drifts away from the bright point, deduced from the east-west direction of the spire drift the east-west direction of the base bipole in about 200 X-ray jets in the polar coronal holes from late 2006 to early 2009.  [A jet's base-bipole east-west direction deduced by the method of Savcheva et al (2009) turns out to the same as that required to fit the observations of Sterling et al (2015) of the production of jets by a filament eruption from the site of the jet-base-edge bright point, even though Savcheva et al (2009) had the wrong reconnecting-field picture for the production of the bright point and for why the jet spire drifts away from the bright point.] Savcheva et al (2009) found that in late 2006 and early 2007, the preferred east-west direction of the jet-base bipoles was that of the active regions of sunspot Cycle 24, the sunspot active regions of which began emerging in early 2008; but that in late 2008 and early 2009 the preferred direction had become that which the active regions of sunspot Cycle 25 will have.  Our global dynamo scenario offers the following explanation for that observation of Savcheva et al (2009):  As Figure 4 indicates, by late 2008 and early 2009 the new (red) high-latitude toroidal field bands for sunspot Cycle 25 had conceivably grown enough for the ephemeral active regions emerging from them in the polar coronal holes to have begun outnumbering the ephemeral active regions emerging there from the toroidal field bands of sunspot Cycle 24.

The analysis of Longcope & Choudhuri (2002) indicates that a whole-active-region $\Omega$ loop remains dynamically connected to its source toroidal field for about four days after emerging, during which time it feels a steady torque from its feet in the source toroidal field band to lose its tilt, that is, to become aligned east west.  Their results indicate that in about four days the convection somehow makes the dispersing $\Omega$ loop dynamically disconnected from the toroidal field band at the bottom: the $\Omega$ loop no longer feels a steady net torque and ceases turning toward east-west alignment.  We suggest that this dynamical disconnection possibly results from the dispersion of the field in the vertical extent of each of the $\Omega$ loop's legs in the convection zone by the free convection, rather than (as Munoz-Jaramillo et al 2010 tacitly assume) by severing the field's connection to the toroidal field band by reconnection.  The mechanism of such dynamical disconnection of an emerged active-region $\Omega$ loop via action of the free convection in the $\Omega$ loop's legs in the convection zone, instead of actual disconnection by reconnection of the two legs with each other, has been simulated by Schussler & Rempel (2005).  Their modeling results have been found to agree with the observed evolutionary decrease in the rotation speed of newly emerged active regions (Svanda et al 2009). So, this dynamical disconnection process by the free convection without severing of the Ω-loop legs by reconnection appears to be a viable way for the results of Longcope & Choudhuri (2002) to be consistent with the dispersing Ω-loop field remaining connected to its toroidal source-field band.

On the other hand, if by about four days after a whole-active-region Ω loop emerges, much of the Ω loop's connection to its source toroidal field band has been severed by reconnection of much of the field in each leg with that in the other leg at some depth in convection zone, then the upper product of that reconnection will be a U loop that has the two upper ends of the U in the two feet of the Ω loop. This reconnection severing would contribute to the Ω loop's dynamical disconnection from its source toroidal field band. Here again, we invoke topological pumping by the free convection in the convection zone to preserve our scenario's essential condition that the Ω loop's dispersing flux remains connected to the Ω loop's source field band. That is, according to our scenario the topological pumping would push the bottom of the U loop to the bottom of the convection zone and, in less than a year, force the field in the bottom of the U loop to reconnect with the Ω loop's source toroidal field band. The result would be that, as its dispersal continues, all of the dispersing flux from the Ω loop remains connected to the Ω loop's source field band, as depicted in each of our four cartoons (Figures 1-4).

Our amended version of Babcock's schematic for the Sun's global dynamo process is only a conceptual heuristic and invokes many assumptions. Even so, we think that our dynamo scenario is plausible enough that it should provoke new observational studies and MHD simulations designed to test its validity. To our knowledge, there is presently no equations-based flux-transport dynamo model that incorporates the entire main new feature of our dynamo scenario, namely that as the field of an emerged large-scale Ω loop is dispersed and transported by surface flows it remains connected to the toroidal field band from which the Ω loop bubbled up through the convection zone, while buoyancy and topological pumping keep the dispersing field roughly vertical in the interior of the convection zone and nearly horizontal along the bottom of the convection zone.

A recent kinematic flux-transport solar dynamo simulation by Yeates & Munoz-Jaramillo (2013) incorporates one of the two aspects of the main new feature of our scenario, namely that the field of emerged active-region Ω loops stays connected to its toroidal source field at the bottom of the convection zone as the Ω-loop fields disperse and their net following-polarity remnants are swept to polar latitudes. The other aspect, that the free convection's topological pumping (acting in concert with buoyancy) is strong enough to keep the poleward-migrating field nearly vertical in the interior of the convection zone and nearly horizontal along the bottom, is not incorporated. In the simulation, the topological pumping in the interior of the convection zone pushes the horizontal component of the poleward-migrating remnant field downward at a speed of ~ 1 m s$^{-1}$, about 10 times slower than the meridional flow of the top of the convection zone sweeps the field's flux at the surface poleward. Consequently, as their Figure 11 shows, in the interior of the convection zone, as the simulation progresses through the maximum phase of sunspot Cycle 23, the following-polarity field that has replaced the opposite-polarity polar cap field of the previous sunspot cycle and the poleward-migrating net following-polarity field at lower latitudes are both far from being nearly radial, in contrast to those fields in our scenario. Instead, in most of the interior of the convection zone in the simulation, those fields have an equatorward-pointing θ component that equals or exceeds the radial component in strength. As their Figure 11 indicates, apparently this results in the next-sunspot-cycle toroidal source-field band being about half as thick verticlly as the convection zone rather than being confined much more closely to the bottom as in our scenario. While the simulation in each polar hemisphere has this quantitative discrepancy with our scenario, we believe that the simulation's production of the next-sunspot-cycle toroidal source-field band poleward of the ongoing-sunspot-cycle source-field band and in the lower convection zone qualitatively demonstrates our scenario. We judge that the same simulation with a factor-of-ten faster speed of downward transport of horizontal field by topological pumping in the interior of the convection zone would mimic our scenario much better. Thus, it appears to us that the solar dynamo simulation of Yeates & Munoz-Jaramillo (2013) is a preliminary positive test of our proposed solar dynamo scenario.

This work was funded by the Heliophysics Division of NASA's Science Mission Directorate through the Living With a Star Targeted Research and Technology Program and the *Hinode* Project. Henk Spruit's Hale Prize Lecture, "How the cycle does and does not work," given 2011 June 14 at the 2011 Meeting of the Solar Physics Division of the American Astronomical Society led RLM to envision the solar dynamo scenario presented here. We also thank Henk Spruit for carefully reading a previous version of the paper and giving helpful feedback. A referee's many insightful comments helped us substantially improve the paper.

REFERENCES

Babcock, H. D. 1959, ApJ, 130, 364.
Babcock, H. W. 1961, ApJ, 133, 572.
Babcock, H. W., & Babcock, H. D. 1955, ApJ, 121, 349.
Charbonneau, P. 2010, Living Rev. Sol. Phys., 7, 3.
Cirtain, J. W., Golub, L., Lundquist, L., et al 2007, Sci, 318, 158.
Fan, Y., Fisher, G. H., & McClymont, A. N. 1994, ApJ, 436, 907.
Hathaway, D. H. 2010, Living Rev. Sol. Phys., 7, 1.
Hathaway, D. H. 2012a, ApJ, 749: L13.
Hathaway, D. H. 2012b, ApJ, 760: 84.
Hathaway, D. H., & Rightmire, L. 2011, ApJ, 729, 80.
Howard, R. 1977, in Illustrated Glossary for Solar and Solar-Terrestrial Physics, ed. A. Bruzek & C. J. Durrant (Dordrecht: Reidel), 7.
Howard, R., & LaBonte, B. J. 1981, SoPh, 74, 131.
Howe, R. 2009, Living Rev. Sol. Phys., 6, 1.
Jackiewicz, J., Serebryanskiy, A., & Kholikov, S. 2015, ApJ, 805:133.
Komm, R. W., Howard, R. F., & Harvey, J. W. 1993, SoPh, 145, 1.
LaBonte, B. J., & Howard, R. 1982, SoPh, 75, 161.
Longcope, D., & Choudhuri, A. R. 2002, SoPh, 205, 63.
McIntosh, S. W., et al 2014, ApJ, 792: 12.
McIntosh, S. W., & Leamon, R. J. 2014, ApJ, 796: L19.

Munoz-Jaramillo, A., Nandy, D., Martens, P.C. H., & Yeats, A. R. 2010, ApJ, 720: L20.

Munoz-Jaramillo, A., Sheeley, N. R., Jr., Zhang, J., & DeLuca, E. E. 2012, ApJ, 753: 146.

Munoz-Jaramillo, A., Dasi-Espuig, M., Balmaceda, L. A., & DeLuca, E. E. 2013, ApJ, 767: L25.

Nelson, N.J., Brown, B. P., Brun, A. S., Miesch, M. S., & Toomre, J. 2013, ApJ, 762: 73.

Nelson, N.J., Brown, B. P., Brun, A. S., Miesch, M. S., & Toomre, J. 2014, SoPh, 289, 441.

Rajaguru, S. P., & Antia, H. M. 2015, ApJ, 813:114.

Savcehva, A., Cirtain, J. W., DeLuca, E. E., & Golub, L. 2009, ApJ, 702: L32.

Schussler, M., & Rempel, M. 2005, A&A, 441, 337.

Spruit, H. C. 2011, in The Sun, the Solar Wind, and the Heliosphere, ed. M. P. Miralles & J. Sanchez Almeida (IAGA Special Sopron Book Series, Vol. 4; Berlin: Springer), 39.

Spruit, H. C., & Roberts, B. 1983, Nature, 304, 401.

Spruit, H. C., & van Ballegooijen, A. A. 1982, A&A, 106, 58.

Sterling, A. C., Moore, R. L., Falconer, D. A., & Adams, M. 2015, Nature, 523, 437.

Svanda, M., Klvana, M., & Sobotka, M. 2009, A&A, 5006, 875.

Topka, K., Moore, R., LaBonte, B. J., & Howard, R. 1982, SoPh, 79, 231.

Wilson, P. R., Altrock, R. C., Harvey, K. L., Martin, S. F., & Snodgrass, H. B. 1988, Nature, 333, 748.

Yeates, A. R., & Munoz-Jaramillo, A. 2013, MNRAS, 436, 3366.

Zhao, J., Bogart, R. S., Kosovichev, A. G., Duvall, T. L., Jr., & Hartlep, T. 2013, ApJ, 774: L29.

Zirin, H. 1988, Astrophysics of the Sun (Cambridge: Cambridge University Press).

Zwaan, C. 1987, ARA&A, 25, 83.